\documentclass[]{spie}  

\usepackage{amsmath,amsfonts,amssymb}
\usepackage{graphicx}
\usepackage{multirow}
\usepackage{comment}
\usepackage[colorlinks=true, allcolors=blue]{hyperref}
\usepackage{amsmath,amsfonts,amssymb}
\usepackage{graphicx}
\usepackage[colorlinks=true, allcolors=blue]{hyperref}
\usepackage{cuted, nccmath}
\usepackage{comment}
\usepackage[utf8x]{inputenc} 

\title{Reliability of deep learning models for anatomical landmark detection: The role of inter-rater variability}

\author[a]{Soorena Salari}
\author[b]{Hassan Rivaz}
\author[a]{Yiming Xiao}
\affil[a]{Computer Science and Software Engineering, Concordia University, Montreal, Canada}
\affil[b]{Electrical and Computer Engineering, Concordia University, Montreal, Canada}

\begin{document} 
\maketitle

\begin{abstract}
Automated detection of anatomical landmarks plays a crucial role in many diagnostic and surgical applications. Progresses in deep learning (DL) methods have resulted in significant performance enhancement in tasks related to anatomical landmark detection. While current research focuses on accurately localizing these landmarks in medical scans, the importance of inter-rater annotation variability in building DL models is often overlooked. Understanding how inter-rater variability impacts the performance and reliability of the resulting DL algorithms, which are crucial for clinical deployment, can inform the improvement of training data construction and boost DL models' outcomes. In this paper, we conducted a thorough study of different annotation-fusion strategies to preserve inter-rater variability in DL models for anatomical landmark detection, aiming to boost the performance and reliability of the resulting algorithms. Additionally, we explored the characteristics and reliability of four metrics, including a novel Weighted Coordinate Variance metric to quantify landmark detection uncertainty/inter-rater variability. Our research highlights the crucial connection between inter-rater variability, DL-models performances, and uncertainty, revealing how different approaches for multi-rater landmark annotation fusion can influence these factors.
\end{abstract}

\keywords{Anatomical landmark detection, Inter-rater variability, Uncertainty, and Deep learning.}

\section{INTRODUCTION}
\label{sec:intro}  
Automated identification of anatomical landmarks plays a vital role in radiological diagnosis, treatment assessment, and surgical planning \cite{proffit2019contemporary,legan1980soft}. Deep learning (DL) methods have recently demonstrated significant advancements in this task \cite{yao2021label,ghesu2016artificial,zhu2021you,tripathi2023unsupervised, salari2023uncertainty,salari2023towards}. It is well acknowledged that applying DL models in safety-critical medical applications demands high accuracy and a thorough understanding of their uncertainty to ensure safety and clinical outcomes. For DL-based anatomical landmark detection, where supervised learning is commonly used, training data with carefully annotated ground truths (GTs) is required, and thus, inter-rater variability is a crucial factor that can affect the algorithms' accuracy and uncertainty. While GT landmarks acquired by fusing annotations from several experts to mitigate potential inconsistencies are commonly utilized if available, they are highly costly, and the best practices to combine multi-rater annotations are still being explored. Inter-rater annotation variability can be caused by multiple factors, such as image noise, experience, rater style, and individual anatomical variations. Such variability in training data will then influence the DL model's performance as part of the uncertainties for the outputs. Therefore, understanding and quantifying the connection between inter-rater variability, accuracy, and uncertainty for DL models in landmark detection can help improve relevant dataset construction and algorithm designs in the future.

While current methods for anatomical landmark detection have yielded promising results and offered uncertainty metrics correlated with detection error \cite{mccouat2022contour,schobs2023bayesian,payer2020uncertainty,melba:2021:014:thaler}, they often employ simplistic and potentially bias-prone approaches to incorporate ground truth annotations from multiple raters, such as averaging coordinates of multi-rater annotations \cite{wang2016benchmark,mccouat2022contour,zhong2019attention,yao2020miss,chen2019cephalometric,zhu2023uod} or selecting only one rater's annotation \cite{schobs2023bayesian,lindner2016fully}. This underscores the importance of exploring more suitable annotation-fusion approaches and a better understanding of the impact of inter-rater variability on the accuracy and reliability of DL-based models for anatomical landmark detection. While this type of investigation in medical AI has been conducted primarily in the realm of medical image segmentation \cite{roshanzamir2023inter,lemay2022label,nichyporuk2022:} and classification \cite{jensen2019improving}, where research has shown that random sampling and Simultaneous Truth and Performance Level Estimation (STAPLE) are useful in preserving inter-rater variability and improving segmentation and classification performance, there has not been any relevant exploration in anatomical landmark detection to the best of our knowledge.

\begin{figure*}[h]
\centering
\includegraphics[scale=0.415]{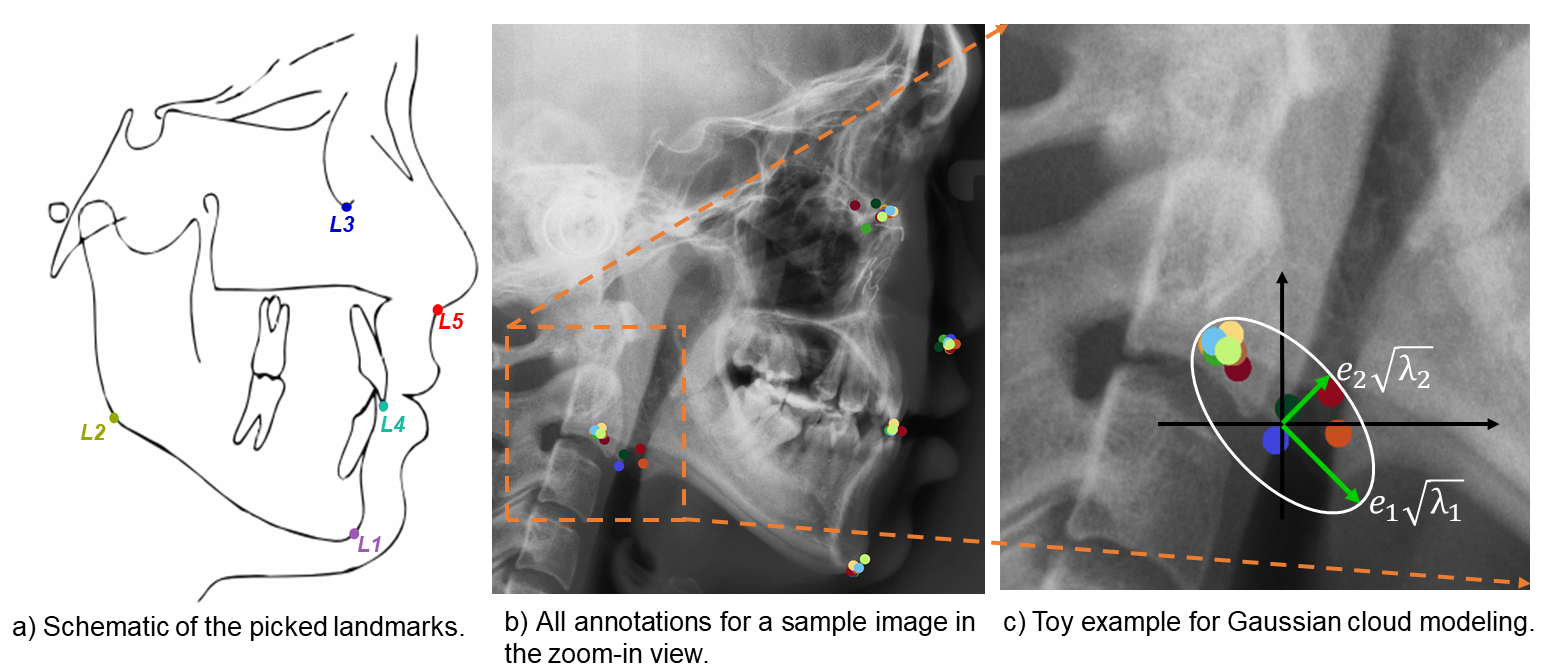}
\caption{ (a) Definition for five target anatomical landmarks. (b) Landmark annotations from 11 raters on an exemplary scan (c) An example for Gaussian-based inter-rater variability modeling (adaptable for uncertainty modeling).}
\label{SampleData}
\end{figure*}

In response to the aforementioned knowledge gap, our study aims to answer two main questions: \textbf{1)} \textit{Which training strategy is optimal for aggregating multiple experts' annotations for DL-based anatomical landmark detection?} Specifically, we compared three different strategies for their ability to preserve inter-rater variability, reduce uncertainty, and enhance performance. \textbf{2)} \textit{What metrics are more effective at capturing and reflecting inter-rater variability and uncertainties in DL models for anatomical landmark detection?} Here, we investigated four different metrics for measuring inter-rater variability and uncertainty in the context of three annotation-fusion strategies. Our work has three main novel contributions. \textbf{First}, we are the first to investigate inter-rater variability in the context of anatomical landmark detection and the impact of different approaches of utilizing multi-rater annotation in this task. \textbf{Second}, we systematically compared the effectiveness of these approaches on accuracy, uncertainty, and inter-rater variability preservation for DL-based anatomical landmark detection methods. \textbf{Third}, we proposed multiple metrics (one novel metric) for landmark detection uncertainty and assessed their ability to reflect model errors and the inter-rater variability. We hope this study will provide useful insights to aid in developing and selecting DL datasets, training methods, and metrics for measuring inter-rater variability and uncertainty for landmark detection. 

\section{Methods and Materials}

\subsection{Dataset and Preprocessing}
\label{DSPP}
Our study utilized the Cephalometric X-ray dataset from the Grand Challenges in ISBI 2015 \cite{wang2016benchmark}. Each cephalometric image, with a 1935$\times$2400 pixels dimension ($\sim$ 0.1$\times$0.1 $mm^{2}$ resolution), contains 19 landmarks marked by both a senior and a junior radiologist. To allow robust inter-rater variability analysis, Franz et al.\cite{melba:2021:014:thaler} selected 100 images from this dataset and tasked nine experienced annotators in medical image analysis to tag 5 representative landmarks out of the original 19 on each image, resulting in 11 annotations in total for each landmark. These landmarks were deliberately chosen to cover a range of difficulty levels and ambiguities, ensuring a wide array of anatomical characteristics (Fig. \ref{SampleData}a). In our investigation, we adopted the selected 100 images along with their 11 sets of annotations for each of the five landmarks. Following prior studies \cite{song2020automatic,payer2019integrating}, we employed downsampled images with the size of 800 $\times$ 640 pixels.

\subsection{Inter-rater Variability}

To quantify inter-rater variability, we employed two types of metrics in our analysis. First, we adopted a metric called coordinate variance ($\textbf{CVar}$), which computes the mean distance of all raters' landmark locations to their averaged coordinates (as the silver ground truth of the real landmark location). This metric is calculated as follows:

\begin{equation}
CVar_k = \frac{1}{N_{\text{Raters}}} \sum_{i=1}^{N_{\text{Raters}}} \|y_{i, k} - \widehat{y}_{k}\|
\end{equation}

\noindent where $CVar_k$ defines the amount of rater variability for the $k_{\text{th}}$ landmark. $N_{\text{Raters}}$ is the total number of raters/samples per single image, $y_{i, k}$ is the annotation of the $i_{\text{th}}$ annotator, and $\widehat{y}_{k}$ is the mean across all annotators. Here, a higher $CVar$ score suggests greater inter-rater variability, reflecting larger discrepancies between individual annotators' annotations and the silver ground truth. 

Additionally, to address the spatial distribution property of different annotations, we modeled the point cloud of annotations as a Gaussian function \cite{melba:2021:014:thaler} and used the related covariance matrix to derive its long and short axis lengths (see Fig. \ref{SampleData}c). Here, we adopted the long axis length, which we refer to as the \textbf{Principal Spatial Variability (PSV)} to help reflect the spatial variability of the point cloud in the longitudinal direction. For an exemplary landmark with $P$ annotations, let $L_{P \times n}$ represent the matrix of annotations where each row is a landmark annotation and $n$ can be up to 3 for 3D landmarks). For calculating the \textit{PSV}, we first calculate the eigenvalues of the covariance matrix of $L$:

\begin{equation}
\{{\lambda_1}, \ldots, {\lambda_n}\} = {\text{eigenvalue}\left(\frac{1}{P}(L-\bar{L})^T(L-\bar{L})\right)}
\end{equation}

\noindent where $\bar{L}$ is the averaged annotation for a landmark location. Then, the \textit{PSV} is defined as the square root of the maximum eigenvalue:

\begin{equation}
\text{PSV} = \sqrt{\lambda_{max}} = \max \{\sqrt{\lambda_1}, \ldots, \sqrt{\lambda_n}\}
\end{equation}

\noindent The higher \textit{PSV} values indicate greater spatial variability within the samples. Furthermore, we also incorporated the \textbf{Anisotropy} of the Gaussian function for each landmark to enrich the dimension of the inter-rater variability measure. It is computed as:

\begin{equation}
\text{Anisotropy} = \frac{\max \{\sqrt{\lambda_1}, \ldots, \sqrt{\lambda_n}\}}{\min \{\sqrt{\lambda_1}, \ldots, \sqrt{\lambda_n}\}+\epsilon} 
\end{equation}

\noindent where $\epsilon$ is a small constant scalar that avoids division by very small values. This metric is very similar to the definition of the Condition Number for the matrices. The more anisotropic the spatial distribution of the underlying point cloud is, the higher the metric goes.

\subsection{Integrating Multi-Rater Annotations and Network Architecture}
\label{IntMultNetArch}

In this study, we employed three distinct annotation-fusion strategies to integrate multi-rater annotations during model training. First, we used the spatial average of all raters' annotations for training each particular landmark, a process we term \textbf{\textit{Averaging}}. Second, during training, we randomly selected the annotation of a single rater from all raters at each iteration, and we refer to this as \textbf{\textit{Random Sampling}}. Lastly, we obtained an ensemble of DL models, each trained on one specific rater's annotations, and then averaged their outputs to generate the final result. We call this method \textbf{\textit{Deep Ensembles}}.

We conducted all experiments using the UNet architecture \cite{ronneberger2015u}, a well-established and efficient architecture in the field of medical image analysis. Notably, UNet has already shown good performance in anatomical landmark detection \cite{yao2020miss,mccouat2022contour,payer2016regressing}. In our study, we employed a UNet model featuring a ResNet-34 encoder pre-trained on natural images. The decoder component of the model comprises five levels of upsampling with channel sizes of 256, 128, 64, 32, and 32. Following each convolution operation, a batch normalization layer and a ReLU activation function are applied. Also, in all decoder levels, we have injected a dropout layer with a probability of 0.5 to avoid overfitting. Additionally, a $1\times1$ convolutional layer is used to reduce the 32 channels in the final layer to 5 channels for 5 target landmarks. Finally, following the success of heatmap-based methods for anatomical landmark detection \cite{melba:2021:014:thaler,mccouat2022contour,payer2020uncertainty,schobs2022uncertainty}, where the location of the landmark is determined by the highest value within the heatmap from the UNet, we have adopted the same strategy for all our experiments.

\subsection{Uncertainty Quantification}

For different annotation-fusion strategies, we devised uncertainty estimation for landmark detection accordingly. For \textit{Averaging} and \textit{Random Sampling}, we incorporated five dropout layers (one per hierarchical level, rate=0.5) in the decoding path of the UNet to produce 20 random samples per landmark based on Monte Carlo (MC) dropout \cite{gal2016dropout} at test time. For \textit{Deep Ensembles}, one sample was collected for each rater-specific model to allow uncertainty measure. Here, we adopted the aforementioned inter-rater variability metrics ($CVar$, $PSV$, and $Anisotropy$) for model uncertainty quantification as well. In addition, considering the values of the heatmap also provide a pseudo-confidence for landmark tagging \cite{schobs2022uncertainty}, we proposed a novel uncertainty metric called the Weighted Coordinate Variance (\textit{$WCVar$}). This metric is calculated as follows:

\begin{equation}
WCVar_k =   \sum_{i=1}^{T} \frac{\frac{1}{\max (h_{i, k})+\epsilon}}{\sum_{i=1}^{T} \frac{1}{ \max (h_{i, k})+\epsilon}}\|y_{i, k} - \widehat{y}_{k}\|
\end{equation}

\noindent where $T$ is the number of predictions/samples per each landmark, where $\epsilon$ is a small constant scalar that avoids division by very small values, $y_{i, k}$ and $h_{i, k}$ are the predicted coordinate and heatmap for the $K_{th}$ landmark in the $i_{th}$ prediction, and $\widehat{y}_{k}$ is the average coordinate of the results among all $T$ predictions. The novelty of \textit{$WCVar$} lies in its ability to integrate coordinate variance with a weighting mechanism that accounts for the spatial distribution of annotations. We hypothesize that this hybrid metric provides a more robust and reliable measure of uncertainty.

\subsection{Implementation and Network Training}
The UNet models for each annotation-fusion strategy were set up according to Section \ref{IntMultNetArch}. For all experiments across the mentioned strategies, we performed a 4-fold cross-validation experiment using the selected 100 scans. We evaluated the model on each fold against the test set and reported the mean values across all four folds for each simulation. During model training, we utilized the AdamW optimizer with a learning rate of $5 \times 10^{-4}$ and a batch size of four for 20 epochs. To minimize the difference between the predicted heatmaps and the ground truths, we employed the negative log-likelihood (NLL) loss function. Additionally, to address the limitations of a small dataset, we implemented extensive data augmentation techniques on the training sets. These techniques included adding random noise, applying random X-Y translations (0 to 10 pixels), scaling intensities randomly, and performing elastic transformations. These augmentations help mitigate overfitting and enhance the model's generalizability. Furthermore, as discussed in Section \ref{DSPP}, our UNet model incorporates a ResNet-34 encoder pre-trained on ImageNet \cite{he2016deep}, which further assists in overcoming the constraints of small datasets.

\section{Experimental Setup and Results}

\subsection{Landmark Detection Accuracy}
To compare the accuracy of different annotation-fusion methods, we employed established evaluation metrics in the literature \cite{wang2015evaluation}. These include the \textbf{Mean Radial Error (MRE)}, which measures the average Euclidean distance between predicted and ground-truth landmarks (in mm), and the \textbf{Success Detection Rate (SDR)} that indicates the percentage of predictions within a set accuracy. Here, we set the SDR thresholds at 2mm, 2.5mm, 3mm, and 4mm.

The MRE and SDR evaluation across all annotation-fusion strategies are shown in Table \ref{Table1} below. With an MRE of 0.68mm and superior SDR at all set thresholds, \textit{Deep Ensembles} outperforms the other two strategies, while the most commonly applied \textit{Averaging} approach ranked last.

\begin{table}[ht]
\centering

\caption{Landmark detection accuracy for different annotation-fusion strategies.}

\begin{tabular}{|c|c|c|c|c|c|c|c|}
\hline 
\multirow[t]{2}{*}{Annotations Fusion} & \multirow[t]{1}{*}{MRE (mm)} & \multicolumn{4}{|c|}{SDR (\%)}  \\
\cline{3-6}
& & $2 \mathrm{~mm}$ & $2.5 \mathrm{~mm}$ & $3 \mathrm{~mm}$ & $4 \mathrm{~mm}$ \\
\hline 
Averaging & 1.42 & 81.8 & 88.2 & 92.2 & 96.4 \\
\hline 
Random Sampling & 1.05 & 89.8 & 93.0 & 95.0 & 98.4\\
\hline 
Deep Ensembles  & 0.68 & 94.8 & 98.4 & 99.0 & 99.8\\
\hline 
\end{tabular}
\label{Table1}
\end{table}

\subsection{Uncertainty and Inter-Rater Variability Quantification} 
For different annotation-fusion methods, we measured their resulting model uncertainty for landmark detection based on $\textit{CVar}$, $\textit{PSV}$, $\textit{Anisotropy}$, and $\textit{WCVar}$, and computed their ``Uncertainty vs. Inter-rater variability" correlations, where both quantities were expressed in matching metrics. Here, when using $\textit{WCVar}$ to quantify inter-rater variability, we assigned each rater's annotation an equal weight for a specific landmark, and the correlations were calculated with bins of 5 landmarks over all landmarks across all scans. The results are shown in Table \ref{Table2} below, where \textit{Deep Ensembles} and \textit{Random Sampling} delivered lower $\textit{CVar}$, $\textit{PSV}$, and $\textit{WCVar}$ than \textit{Averaging}. However, all strategies produced similar $\textit{Anisotropy}$ values. As for the correlations, which reflect the model's ability to preserve inter-rater variability, \textit{Deep Ensembles} and \textit{Random Sampling} surpassed \textit{Averaging}.

\begin{table*}[htbp]
\centering
\renewcommand{\arraystretch}{2}
\caption{Mean values of different uncertainty metrics for all annotation-fusion strategies and their ``Uncertainty vs. Inter-rater variability" Pearson correlation coefficients, where both quantities were expressed in matching metrics.}
\resizebox{\textwidth}{!}{
\begin{tabular}{|c|c|c|c|c|c|c|c|c|}
\hline 
\multirow{2}{*}{Annotation-Fusion} & \multicolumn{4}{|c|}{Mean Uncertainty metrics} & \multicolumn{4}{|c|}{``Uncertainty vs. Inter-rater Variability" Correlation}  \\
\cline{2-9}
& CVar (mm) & PSV (mm) & Anisotropy & WCVar (mm) & CVar & PSV  &  Anisotropy & WCVar  \\
\hline 
Averaging & 2.36 & 2.76 & 2.50 & 2.33 & 0.05 & -0.006 & 0.15 & 0.08 \\
\hline 
Random Sampling & 1.25 & 1.33 & 2.51 & 1.34 & 0.50 & 0.46 & 0.74 & 0.53 \\
\hline 
Deep Ensembles  &  1.02 & 1.10 & 2.76 & 1.05 & 0.82 & 0.76 & 0.25 & 0.85 \\
\hline 
\end{tabular}
}
\label{Table2}

\end{table*}

\subsection{Reliability of Different Uncertainty Metrics} 
Reliable uncertainty metrics of DL models should possess a strong correlation with the errors. To compare the reliability of our proposed uncertainty metrics, we computed their correlations against the landmark detection errors (i.e., Euclidean distance between the predicted and GT landmarks) under three annotation-fusion strategies for all landmarks across all scans. The results are listed in Table \ref{Table3} below, which shows that $\textit{CVar}$ and $\textit{WCVar}$ generally outperform the other metrics. Moreover, \textit{Averaging} yielded superior results compared to the other annotation-fusion strategies, which have similar reliability.

\begin{table}[htbp]
\centering
\caption{Pearson correlation coefficients of the proposed uncertainty metrics with the landmark detection error for different annotation-fusion strategies.}

\begin{tabular}{|c|c|c|c|c|}
\hline 
\multirow{2}{*}{Annotation-Fusion} & \multicolumn{4}{|c|}{``Uncertainty vs. Landmark Detection Error" Correlation} \\
\cline{2-5}
& CVar & PSV & Anisotropy & WCVar \\
\hline 
Averaging & 0.93 & 0.85 & 0.47 & 0.94\\
\hline 
Random Sampling & 0.65 & 0.68 & 0.01 & 0.69 \\
\hline 
Deep Ensembles  &  0.68 & 0.69 & -0.01 & 0.70 \\
\hline 
\end{tabular}

\label{Table3}
\end{table}

\section{Discussion}

Due to the unique problem setting, fusing multi-rater annotation for landmark points and the related uncertainty/inter-rater variability quantification required distinct approaches from segmentation and classification \cite{roshanzamir2023inter,lemay2022label,jensen2019improving}. Therefore, we have proposed our metrics and experiments accordingly. Based on the results from Tables \ref{Table1} and \ref{Table2}, for annotation-fusion strategies, we found the benefits of \textit{Random Sampling} and \textit{Deep Ensembles} over the simplistic but commonly used point \textit{Averaging} for improving the DL model's accuracy and confidence. These performance boosts were accompanied by better preservation of inter-rater variability, and \textit{Deep Ensembles} presented a slight edge over \textit{Random Sampling}. In general, these observations echoed those from medical image segmentation and classification \cite{roshanzamir2023inter,lemay2022label,jensen2019improving}. Despite the advantage of \textit{Deep Ensembles} for DL model performance, with the consideration of its high computational cost, \textit{Random Sampling} can be a good alternative choice. To adapt to the specific problem of landmark detection, we leveraged four different metrics to quantify and better characterize model uncertainty and inter-rater variability. These metrics capture the tightness, anisotropy, and extent of the target point cloud. Notably, Schobs et al. \cite{schobs2022uncertainty} recently proposed to use the inverse of the max value in a heatmap from a landmark detection UNet to represent the uncertainty but also found it to be susceptible to the distribution of the point cloud. To mitigate this, we proposed a novel \textit{Weighted Coordinate Variance (WCVar)} to combine this notion with the distance-based $\textit{CVar}$ metric. Based on the correlation studies in Tables \ref{Table2} and \ref{Table3}, we observed different behaviors of these metrics for assessing inter-rater variability preservation and reliability of uncertainty quantification. While all metrics were able to reveal similar general trends for inter-rater variability preservation, the newly proposed $\textit{WCVar}$ has demonstrated its superior reliability for uncertainty quantification, with a stronger correlation to detection errors. One interesting observation is the stronger ``Uncertainty vs. Error" correlation for \textit{Averaging} in Table \ref{Table3}. We suspect that this is due to the use of averaged annotations as silver GT for evaluation, and this calls for more sophisticated techniques for annotation fusion in landmark detection. For our study, we relied on MC dropout and model ensembles, which approximate the true Bayesian neural networks (BNNs) and are model-agnostic for epistemic uncertainty estimation. While aleatoric uncertainty also contributes to the total uncertainty and inter-rater variability, the relevant quantification still requires further investigation for landmark detection. Due to limited data size, we only performed our experiments based on a UNet. As DL model architectures may also affect accuracy, inter-rater variability, and uncertainty \cite{roshanzamir2023inter}, we will explore this venue in the future. Finally, our current case study employed 2D X-ray images. To help generalize the insights, it is beneficial to conduct similar studies in 3D landmarks in MRI and CT, but the experiments can be limited by the availability of relevant datasets.

\section{Conclusions}

This paper intends to fill the knowledge gap regarding inter-rater variability in DL-based anatomical landmark detection. Specifically, we examined three different strategies (\textit{Averaging}, \textit{Random Sampling}, and \textit{Deep Ensembles}) of multi-rater annotation-fusion for the interplay of inter-rater variability preservation, model accuracy, and epistemic uncertainty. Additionally, we explored the reliability of four different metrics (including the novel \textit{WCVar}) to quantify uncertainty and inter-rater variability for landmark detection. Our case study revealed that annotation \textit{Averaging}, which is frequently used, can negatively impact the performance and uncertainty of DL models. On the other hand, we showed that \textit{Random Sampling} and \textit{Deep Ensembles} could boost the preservation of inter-rater variability and accuracy of the model while lowering the epistemic uncertainty.

\section{Acknowledgment}
We acknowledge the support of the Natural Sciences and Engineering Research Council of Canada (NSERC).

\bibliography{main} 

\begin{thebibliography}{10}

\bibitem{proffit2019contemporary}
Proffit, W.~R., Fields, H., Msd, D.~M., Larson, B., and Sarver, D.~M.,  [{\em Contemporary Orthodontics, 6e: South Asia Edition-E-Book}{\nolinebreak\hspace{0.1em}]}, Elsevier India (2019).

\bibitem{legan1980soft}
Legan, H.~L. and Burstone, C.~J., ``Soft tissue cephalometric analysis for orthognathic surgery.,'' {\em Journal of Oral Surgery (American Dental Association: 1965)}~{\bf 38}(10),  744--751 (1980).

\bibitem{yao2021label}
Yao, Q., Xiao, L., Liu, P., and Zhou, S.~K., ``Label-free segmentation of {COVID}-19 lesions in lung {CT},'' {\em IEEE Transactions on Medical Imaging}~{\bf 40}(10),  2808--2819 (2021).

\bibitem{ghesu2016artificial}
Ghesu, F.~C., Georgescu, B., Mansi, T., Neumann, D., Hornegger, J., and Comaniciu, D., ``An artificial agent for anatomical landmark detection in medical images,'' in [{\em International Conference on Medical Image Computing and Computer-Assisted Intervention}{\nolinebreak\hspace{0.1em}]},   229--237, Springer (2016).

\bibitem{zhu2021you}
Zhu, H., Yao, Q., Xiao, L., and Zhou, S.~K., ``You only learn once: {U}niversal anatomical landmark detection,'' in [{\em International Conference on Medical Image Computing and Computer-Assisted Intervention}{\nolinebreak\hspace{0.1em}]},   85--95, Springer (2021).

\bibitem{tripathi2023unsupervised}
Tripathi, A., Panicker, M.~R., Hareendranathan, A.~R., Jaremko, J., Chen, Y.~T., Narayan, K.~V., and Kesavadas, C., ``Unsupervised landmark detection and classification of lung infection using transporter neural networks,'' {\em Computers in Biology and Medicine}~{\bf 152},  106345 (2023).

\bibitem{salari2023uncertainty}
Salari, S., Rasoulian, A., Battie, M., Fortin, M., Rivaz, H., and Xiao, Y., ``Uncertainty-aware transformer model for anatomical landmark detection in paraspinal muscle {MRI}s,'' in [{\em Medical Imaging 2023: Image Processing}{\nolinebreak\hspace{0.1em}]},   {\bf 12464},  246--252, SPIE (2023).

\bibitem{salari2023towards}
Salari, S., Rasoulian, A., Rivaz, H., and Xiao, Y., ``Towards multi-modal anatomical landmark detection for ultrasound-guided brain tumor resection with contrastive learning,'' in [{\em International Conference on Medical Image Computing and Computer-Assisted Intervention}{\nolinebreak\hspace{0.1em}]},   668--678, Springer (2023).

\bibitem{mccouat2022contour}
McCouat, J. and Voiculescu, I., ``Contour-hugging heatmaps for landmark detection,'' in [{\em Proceedings of the IEEE/CVF Conference on Computer Vision and Pattern Recognition}{\nolinebreak\hspace{0.1em}]},   20597--20605 (2022).

\bibitem{schobs2023bayesian}
Schobs, L., McDonald, T.~M., and Lu, H., ``Bayesian uncertainty estimation in landmark localization using convolutional gaussian processes,'' in [{\em International Workshop on Uncertainty for Safe Utilization of Machine Learning in Medical Imaging}{\nolinebreak\hspace{0.1em}]},   22--31, Springer (2023).

\bibitem{payer2020uncertainty}
Payer, C., Urschler, M., Bischof, H., and {\v{S}}tern, D., ``Uncertainty estimation in landmark localization based on gaussian heatmaps,'' in [{\em International Workshop on Uncertainty for Safe Utilization of Machine Learning in Medical Imaging}{\nolinebreak\hspace{0.1em}]},   42--51, Springer (2020).

\bibitem{melba:2021:014:thaler}
Thaler, F., Payer, C., Urschler, M., and Štern, D., ``Modeling annotation uncertainty with gaussian heatmaps in landmark localization,'' {\em Machine Learning for Biomedical Imaging}~{\bf 1},  1--27 (2021).

\bibitem{wang2016benchmark}
Wang, C.-W., Huang, C.-T., Lee, J.-H., Li, C.-H., Chang, S.-W., Siao, M.-J., Lai, T.-M., Ibragimov, B., Vrtovec, T., Ronneberger, O., et~al., ``A benchmark for comparison of dental radiography analysis algorithms,'' {\em Medical Image Analysis}~{\bf 31},  63--76 (2016).

\bibitem{zhong2019attention}
Zhong, Z., Li, J., Zhang, Z., Jiao, Z., and Gao, X., ``An attention-guided deep regression model for landmark detection in cephalograms,'' in [{\em International Conference on Medical Image Computing and Computer-Assisted Intervention}{\nolinebreak\hspace{0.1em}]},   540--548, Springer (2019).

\bibitem{yao2020miss}
Yao, Q., He, Z., Han, H., and Zhou, S.~K., ``Miss the point: Targeted adversarial attack on multiple landmark detection,'' in [{\em International Conference on Medical Image Computing and Computer-Assisted Intervention}{\nolinebreak\hspace{0.1em}]},   692--702, Springer (2020).

\bibitem{chen2019cephalometric}
Chen, R., Ma, Y., Chen, N., Lee, D., and Wang, W., ``Cephalometric landmark detection by attentive feature pyramid fusion and regression-voting,'' in [{\em International Conference on Medical Image Computing and Computer-Assisted Intervention}{\nolinebreak\hspace{0.1em}]},   873--881, Springer (2019).

\bibitem{zhu2023uod}
Zhu, H., Quan, Q., Yao, Q., Liu, Z., and Zhou, S.~K., ``{UOD}: Universal one-shot detection of anatomical landmarks,'' in [{\em International Conference on Medical Image Computing and Computer-Assisted Intervention}{\nolinebreak\hspace{0.1em}]},   24--34, Springer (2023).

\bibitem{lindner2016fully}
Lindner, C., Wang, C.-W., Huang, C.-T., Li, C.-H., Chang, S.-W., and Cootes, T.~F., ``Fully automatic system for accurate localisation and analysis of cephalometric landmarks in lateral cephalograms,'' {\em Scientific Reports}~{\bf 6}(1),  33581 (2016).

\bibitem{roshanzamir2023inter}
Roshanzamir, P., Rivaz, H., Ahn, J., Mirza, H., Naghdi, N., Anstruther, M., Batti{\'e}, M.~C., Fortin, M., and Xiao, Y., ``How inter-rater variability relates to aleatoric and epistemic uncertainty: a case study with deep learning-based paraspinal muscle segmentation,'' in [{\em International Workshop on Uncertainty for Safe Utilization of Machine Learning in Medical Imaging}{\nolinebreak\hspace{0.1em}]},   74--83, Springer (2023).

\bibitem{lemay2022label}
Lemay, A., Gros, C., Naga~Karthik, E., and Cohen-Adad, J., ``Label fusion and training methods for reliable representation of inter-rater uncertainty,'' {\em Machine Learning for Biomedical Imaging}~{\bf 1},  1--27 (2022).

\bibitem{nichyporuk2022:}
Nichyporuk, B., Cardinell, J., Szeto, J., Mehta, R., Falet, J.-P., Arnold, D.~L., Tsaftaris, S.~A., and Arbel, T., ``Rethinking generalization: The impact of annotation style on medical image segmentation,'' {\em Machine Learning for Biomedical Imaging}~{\bf 1},  1--37 (2022).

\bibitem{jensen2019improving}
Jensen, M.~H., J{\o}rgensen, D.~R., Jalaboi, R., Hansen, M.~E., and Olsen, M.~A., ``Improving uncertainty estimation in convolutional neural networks using inter-rater agreement,'' in [{\em International Conference on Medical Image Computing and Computer-Assisted Intervention}{\nolinebreak\hspace{0.1em}]},   540--548, Springer (2019).

\bibitem{song2020automatic}
Song, Y., Qiao, X., Iwamoto, Y., and Chen, Y.-w., ``Automatic cephalometric landmark detection on {X}-ray images using a deep-learning method,'' {\em Applied Sciences}~{\bf 10}(7),  2547 (2020).

\bibitem{payer2019integrating}
Payer, C., {\v{S}}tern, D., Bischof, H., and Urschler, M., ``Integrating spatial configuration into heatmap regression based {CNN}s for landmark localization,'' {\em Medical Image Analysis}~{\bf 54},  207--219 (2019).

\bibitem{ronneberger2015u}
Ronneberger, O., Fischer, P., and Brox, T., ``U-net: {C}onvolutional networks for biomedical image segmentation,'' in [{\em International Conference on Medical Image Computing and Computer-Assisted Intervention}{\nolinebreak\hspace{0.1em}]},   234--241, Springer (2015).

\bibitem{payer2016regressing}
Payer, C., {\v{S}}tern, D., Bischof, H., and Urschler, M., ``Regressing heatmaps for multiple landmark localization using {CNN}s,'' in [{\em International Conference on Medical Image Computing and Computer-assisted Intervention}{\nolinebreak\hspace{0.1em}]},   230--238, Springer (2016).

\bibitem{schobs2022uncertainty}
Schobs, L.~A., Swift, A.~J., and Lu, H., ``Uncertainty estimation for heatmap-based landmark localization,'' {\em IEEE Transactions on Medical Imaging}~{\bf 42}(4),  1021--1034 (2022).

\bibitem{gal2016dropout}
Gal, Y. and Ghahramani, Z., ``Dropout as a bayesian approximation: Representing model uncertainty in deep learning,'' in [{\em International {C}onference on {M}achine {L}earning}{\nolinebreak\hspace{0.1em}]},   1050--1059, PMLR (2016).

\bibitem{he2016deep}
He, K., Zhang, X., Ren, S., and Sun, J., ``Deep residual learning for image recognition,'' in [{\em Proceedings of the IEEE/CVF Conference on Computer Vision and Pattern Recognition}{\nolinebreak\hspace{0.1em}]},   770--778 (2016).

\bibitem{wang2015evaluation}
Wang, C.-W., Huang, C.-T., Hsieh, M.-C., Li, C.-H., Chang, S.-W., Li, W.-C., Vandaele, R., Mar{\'e}e, R., Jodogne, S., Geurts, P., et~al., ``Evaluation and comparison of anatomical landmark detection methods for cephalometric {X}-ray images: a grand challenge,'' {\em IEEE Transactions on Medical Imaging}~{\bf 34}(9),  1890--1900 (2015).

\end{thebibliography}
\bibliographystyle{spiebib} 

\end{document}